\documentclass[12pt]{article}
\usepackage{graphicx}
\begin{document}
\centerline {\bf Election results and the Sznajd model on Barabasi network}

\bigskip
A. T. Bernardes$^{1,2}$, D. Stauffer$^{2}$ and J. Kert\'esz$^{3}$

\bigskip

\noindent
$^1$ Departamento de F\'{\i}sica, Universidade Federal de Ouro Preto, Campus 
do Morro do Cruzeiro, 35400-000, Ouro Preto-MG, Brazil 

\medskip
\noindent
$^2$ Institute for Theoretical Physics, Cologne University, 
D-50923 K\"oln, Euroland

\medskip
\noindent
$^3$Department of Theoretical Physics, Budapest University of
Technology and Economics, Budafoki \'ut 8, H-1111, Budapest, Hungary

\medskip
\noindent
e-mails: atb@iceb.ufop.br, stauffer@thp.uni-koeln.de, kertesz@phy.bme.hu
\bigskip

\begin{abstract}
The network of Barabasi and Albert, a preferential
growth model where a new node is linked to the old ones with a
probability proportional to their connectivity, is applied to Brazilian election
results. The application of the Sznajd rule, that only agreeing pairs of people
can convince their neighbours, gives a vote distribution in good agreement 
with reality.

\end{abstract}


\section{Introduction}

Nowadays, it has been a matter of increasing interest \cite{exotic} to apply
the fundamentals of the theories of complex systems 
in many different disciplines,
not only in physical sciences, but even in social sciences, from
economy to education \cite{albano} or sociology. The main point is that 
social systems, like natural ones, are constituted of great
number of individuals, which - generally - have local interactions
between them. Sometimes, social networks behaviour can be determined also
by the action of external actors, which might
be mimicked by external fields in our model.

Elections are processes where many individuals interact between 
them. It is a dynamical convincing process, where we have at the same time
the interaction between neighbours and external influence (political
advertising, campaigns etc). In Brazil, in proportional elections
(deputies or city councillors) the voters vote directly for the candidates
and not for the parties. They can vote for a party, but it is
not frequent. Some elections occur with a large number of voters:
In some states or in the largest cities one has a number
of voters in the order of magnitudes of millions or tens of millions.
So, these elections are a social phenomenon which presents the basic
characteristics of complex systems. One of these features is that
they are scale-free phenomena. This feature has been observed
by Costa-Filho {\it et al.} \cite{soares}, who showed that the distribution
of the number of votes obtained by different candidates
for the 1998 elections in Brazil
follow a power law distribution, with exponent $\simeq -1.0$. The
same result can be obtained for the whole country or for other proportional
elections, which shows that this is a very robust result. 

Taking into account that elections are processes where 
a vote is supposed to be obtained as a result of convincing arguments, 
it can be compared
with a physical process of clustering. However, contrary to the
results obtained by Costa-Filho {\it et al}, usual models of formation of 
clusters (as percolation, for instance \cite{perc}) may give exponents
$\simeq -2.0$ (square lattice) 
for the numbers of clusters as a function of the cluster size.

Recently, we have introduced \cite{bernardes} a model for proportional
elections. Our model is based on the Sznajd model proposed
to simulate the process of formation of opinion. However,
different from other models \cite{holyst}, where the influence flows inward 
from the border to the center (like in the majority games,
where the site in the middle takes the state of the majority of
neighbouring sites), in the Sznajd model
\cite{sznajd1,sznajd2,sznajd3,sznajd4} one has an outward flow of influence. 
It thus differs from e.g. bootstrap percolation \cite{adler} or other cellular
automata (for a computational review see \cite{automata}) where the site in
the center behaves according to a rule determined by its neighbours. 
Nevertheless, the dynamics of the Sznajd model is quite similar (except for 
isolated sites) to that of spinodal decomposition of the Ising model \cite{liv}
at low temperatures, as was shown in \cite{bernardes}: 
Starting from a random distribution, large domains form
where nearly all sites have the same state. Finally, one domain will cover the 
whole lattice. Thus to get the desired results we will look at intermediate
times when there are still many different domains or correlated sites in the
system. 

In the Sznajd model, small sets of people influence the opinions of their 
nearest neighbours if and only if all people within the original set agree.
On a chain, this set is a bond with
two people at its ends \cite{sznajd1,sznajd3}. On the square lattice with
\cite{sznajd4} or without \cite{sznajd2} disorder, it can be such a bond or a 
plaquette of four neighbouring spins (people). This plaquette rule is called
rule Ia in \cite{sznajd2} and the bond rule, used in the present paper, is
called rule IIa in \cite{sznajd2}. 
Thus if all four plaquette members or two bond
members share the same opinion, they convince their neighbours of this opinion; 
otherwise the neighbours remain unchanged.

However, as shown by Barabasi and Albert recently \cite{barabasi}, social
relations must be represented by networks instead of lattices.
Networks of interactions (www, author's collaboration in scientific
papers, actor's collaborations in films) show the common feature
of scale-free behaviour. In order to represent this main
feature, Barabasi and Albert introduced a model for evolving networks.
Numerous papers used this model for a variety of purposes, e.g. \cite{review}.
Starting with few nodes (which may represent actors, authors, web sites)
connected to each other, more and more nodes are added to the network,
each node connecting to an already connected one, with the probability
to connect to a node being proportional to the number of previous nodes
which are already connected to it.

In Brazil the voting process is much
more based on the relation between candidate/voter than on the
parties. Thus, our idealized version of the voting process can
omit the role of the parties. Another aspect is that it was
clear for us that the candidates do not start with the same
social weight. This determines the result of the elections, since
candidates with more social visibility or better conditions
to campaign are more likely to be elected. So, we have
to introduce some differentiation between the candidates in the
beginning of the simulation.

In this paper, we performed simulations on a three-dimensional
version of the previous one-dimensional \cite{sznajd1} and two-dimensional
\cite{sznajd2} Sznajd model. We combine it with 
a network model 
for elections based on the model of Bernardes et al \cite{bernardes}. 
Unlike this first version \cite{bernardes} and its three-dimensional variant
where a probability to convince had to be introduced, eq.(1) below, in order
to produce some differentiation between the candidates, with the Barabasi
network the same result is 
obtained from the combination of the different number of neighbours of the nodes
without this probability. 

In the next section we present the models we have used, followed by the results.
Both of these models use networks connecting the voters; one network is a 
simple cubic lattice, the other a Barabasi network. After that, we conclude.

\section{Models and Results}

We have simulated two models in the present work. The first
one is a 3d version of that simulated previously \cite{bernardes}. 
The second is a Barabasi network version. 

\subsection{Simple cubic lattice}

In this work, we used a modified version of the Sznajd model (rule IIa in 
\cite{sznajd2}): A pair of neighbours in agreement convinces its 
ten nearest neighbours to the same opinion. A cubic lattice of size 
$L \times L \times L$ represents the set of voters. A number 
$N_{tot}$ of candidates, $N_{tot} \ll L^3$, is fixed in the beginning of the 
simulation.  The value $n = 1,2, \dots, N_{tot}$ of a site $S$ on the lattice 
will represent that this voter prefers that candidate $n$. The model
has two different stages: First, we  produce the initial condition and,
after that, we perform the simulation of the electoral
campaign (only voters can influence other voters, {\it a la} Sznajd).
As in real elections, we do not wait for a kind of equilibrium state, but count 
the votes at some intermediate time. Basically what we are doing is the 
analysis during the transient time. As in real
elections, the candidates have different initial 
chances of being voted for (representing more money for electoral
campaigns, more initial visibility etc.). This is modelled by
a probability $P_c$ of convincing, calculated from the label $n$
of the candidate

$$ P_c= (n/N_{tot})^2 \quad. \eqno (1)$$

\noindent It means that the higher is the label $n$ of a candidate,
the higher is the probability of convincing a voter.

In the first stage, we started with all the sites with value zero, meaning
that there are no committed voters. 
Then, we visit all the sites exactly once, in random order. For each
visit, we try to convince the voter to adopt a candidate, chosen
at random. A random number $r$ is generated and compared with $P_c$.
If $r \leq P_c$ the candidate is accepted by that voter. 
If the candidate convinces the voter, this voter tries to convince
the neighbouring sites. Once again, we throw the dice and compare
a new random number with $P_c$. If successful, $r \leq P_c$, the voter will
try to convince the neighbourhood as follows: We check all the six neighbouring
sites; for each that has the same value of the candidate chosen
before, all the ten neighbouring sites of this bond of two sites
will assume the same value (as in the usual Sznajd prescription).
If nobody has chosen the same candidate, only the originally selected
voter is committed to this candidate.

In the second stage, a usual Sznajd process is performed without using the
complication from the probability $P_c$. (We thus assume all voters to be 
equal and restrict the probability $P_c$ to describe the convincing power of 
the candidates only.) We go
to random sites on the lattice. A neighbouring site is chosen
at random and we check if the two sites have the same value (they
prefer the same candidate). In that case, all the ten neighbours
change to vote in that candidate.

\begin{figure}[hbt]
\begin{center}
\includegraphics[angle=-90,scale=0.5]{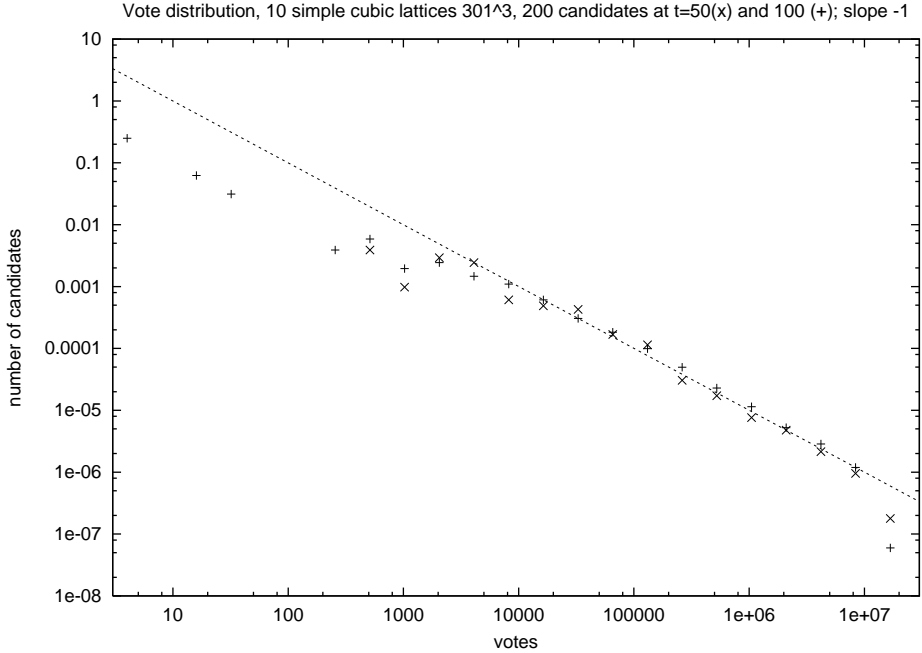}
\end{center}
\caption{
Distribution $N(v)$ of the number of candidates getting $v$ votes each
on the simple cubic lattice, after 50 and 100 iterations. Election of 200
candidates by 27 million voters.
}
\end{figure}    

Figure 1 shows, just as in real life \cite{soares}, deviations from a simple
power law for both very large numbers of votes and very small numbers. In
between, however, the simulations are compatible with the hyperbolic law
$$ N(v) \propto 1/v \eqno (2)$$
observed in reality
for the number $N$ of candidates having $v$ votes each. (Here and in Fig.2 below
the bin size for $v$ increases by a factor 2 for each consecutive bin.) 

[In \cite{bernardes}, for the square lattice
and assumption (1), {\it two} exponents are fitted onto the data: One for the
distribution after the first stage, and one for the distribution during 
the second change. When assumption (1) is generalized from $\propto n^2 $
to $\propto n^x$, then the first exponent depends strongly on $x$ (which can be 
explained by a simple analytical relation between this exponent and $x$), while
the second exponent depends much less on $x$.]

\subsection{Network version}

In this model, we first create a network of interacting nodes
by using the basic Barabasi-Albert prescription. We  fix
an initial number of nodes, each one connected to the others.
In the present work, the minimum number of connections of a node is 
$m=5$. So, in the beginning
of the simulation we have 6 nodes, in order that each one can
be connected to the 5 others. After that, more 
and more nodes are added to the network. A new node has a probability to 
be connected to a previous node proportional
to the number of nodes that are already connected to this previous node.
Thus the growth probability at any existing node
is proportional to the number of nodes already connected to it.
We will no longer need assumption (1) and have replaced this assumption and
the regular lattice by the Barabasi network without such an assumption.

After preparing the network, we start with the election process, which is now
different from that in the previous three-dimensional model. The first step
is the distribution of candidates. Again, the state of a node,
that means, the value $n$ of a node on the network, represents that this
voter has given the preference to that candidate $n$. 
Thousand candidates are distributed at random, disregarding the
number of connections of a node, i.e. we pick a node at random from the half
million nodes to which we let the network grow, and then 
a candidate at random. Now, the campaign starts. At each time step
we visit all the nodes. For each node, we have the following process:

\begin{itemize}

\item If a node $i$ has already selected preference for a candidate, we choose
a connected node $j$ at random. If node $i$ has no candidate ($n=0$),
we go to another randomly selected node.

\item If node $j$ has the same candidate as node $i$, they try to convince all
the nodes connected with them. The probability to convince others for each of
the two nodes is now inversely proportional to the time-independent 
numbers of nodes connected with it, meaning
that each node convinces - on average -  one other node at each process.

\item If node $j$ has no candidate, node $i$ tries to convince it
to accept its own candidate, with the same probability as described above.

\item If node $j$ has a different candidate from node $i$, we skip
to another node $i$.

\end{itemize}

Again, as described above for the 3d version, we do not wait for
a equilibrium state. It is important to mention that, different
from a square lattice, where an equilibrium state is reached
in a time proportional to the number of sites, in the network
it is reached rapidly, after about $10^2$ iterations. 
In both cases, in final equilibrium all the sites have the same state.

Figure 2 shows that again except for the smallest and the largest numbers $v$
of votes, the hyperbolic law (2) is obeyed well at intermediate times $t=40$.

\section{Summary}
Whether we simulate the election process on a square lattice, a simple cubic 
lattice or a Barabasi network, we recover the same hyperbolic law as found in
real elections. Our simulations on a Barabasi network have the advantage that 
we no longer need assumption (1) for the purpose of getting a realistic
vote distribution with decay exponent 1 in the center.

Our study has shown that the hyperbolic law observed empirically is a
rather robust consequence of our modifid Sznajd model, since we found it first
on the square lattice \cite{bernardes} and now on both the cubic lattice
and the Barabasi network.
Either we use a regular lattice and assumption (1), or we
use the Barabasi network without assumption (1); the final results are similar.
The fact that the hyperbolic law is observable on the Barabasi
network, which is a more realistic model of social interactions than
the lattices, provides evidence that the Sznajd model may well capture
important aspects of the voting mechanism. The advantage of using the
Barabasi network instead of regular lattices is not only that it is more
realistic but also that we can drop assumption (1) which is a kind of
fine tuning the system to criticality. Of course, the Barabasi model,
and the related assumptions are also rules (as we have rules when
constructing a lattice too) but the difference is similar to what we
have for usual and self-organized criticality: Assumption (1) puts in some
exponent at the beginning, through $(n/N_{tot})^2$,
on which the final exponent depends somewhat (see end of section 2.1), while 
the Barabasi growth process leads by itself to a power-law distribution of the
number of connections for a node, and combined with the Sznajd model gives
the desired vote distribution with its intermediate power law.

Moreover, rule (1) was introduced ad hoc to explain the election results, while
the rules of \cite{barabasi} were stated before, independent of the present 
application. We are not aware of other voter models\cite{hinrichsen}
explaining the hyperbolic law found empirically in \cite{soares}.
\bigskip

\noindent {\bf Acknowledgments}: We thank Ana Proykova
for suggesting the 3d simulation.  ATB acknowledges the hospitality 
of the Institute for Theoretical Physics from the University
of Cologne. This work was partially supported by the
Brazilian Agencies CNPq, FINEP and by the Hungarian OTKA T029985.

\begin{figure}[hbt]
\begin{center}
\includegraphics[angle=-90,scale=0.50]{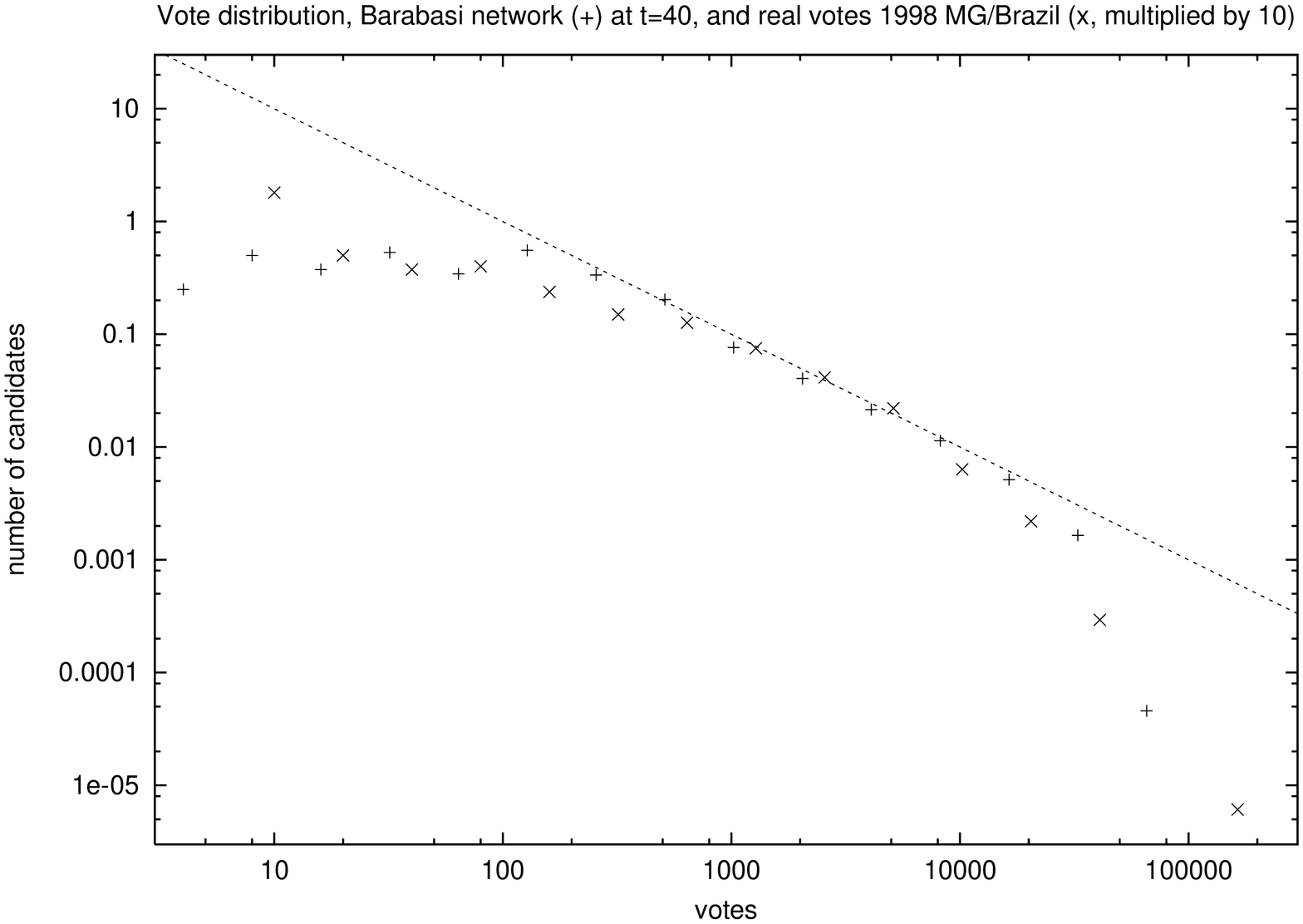}
\end{center}
\caption{
Distribution $N(v)$ for half a million nodes on the Barabasi network,
where each previously added node bonds to five previously added nodes.  Election
of 1000 candidates (+). The number of votes in real elections ($\times$:
state of Minas Gerais in Brazil 1998) is multiplied by ten for better
comparison.
}
\end{figure}    
\end{document}